\documentclass{article}
\usepackage{geometry,amsmath,amssymb,graphicx,enumerate,calc,color,epsfig}    %%%ifthen,capt-of}
\newcommand{\emdash}{---}
\newcommand{\tmem}[1]{{\em #1\/}}
\newcommand{\tmname}[1]{\textsc{#1}}
\newcommand{\tmop}[1]{\ensuremath{\operatorname{#1}}}
\newcommand{\tmsamp}[1]{\textsf{#1}}
\newcommand{\tmtextbf}[1]{{\bfseries{#1}}}
\newcommand{\tmtextit}[1]{{\itshape{#1}}}
\newcommand{\tmtextsc}[1]{{\scshape{#1}}}
\newcommand{\tmtexttt}[1]{{\ttfamily{#1}}}
\newenvironment{enumeratenumeric}{\begin{enumerate}[1.] }{\end{enumerate}}
\definecolor{grey}{rgb}{0.75,0.75,0.75}
\definecolor{orange}{rgb}{1.0,0.5,0.5}
\definecolor{brown}{rgb}{0.5,0.25,0.0}
\definecolor{pink}{rgb}{1.0,0.5,0.5}
\newcommand{\tmfloatcontents}{}
\newlength{\tmfloatwidth}
\newcommand{\tmfloat}[5]{
  \renewcommand{\tmfloatcontents}{#4}
  \setlength{\tmfloatwidth}{\widthof{\tmfloatcontents}+1in}
  \ifthenelse{\equal{#2}{small}}
    {\ifthenelse{\lengthtest{\tmfloatwidth > \linewidth}}
      {\setlength{\tmfloatwidth}{\linewidth}}{}}
    {\setlength{\tmfloatwidth}{\linewidth}}
  \begin{minipage}[#1]{\tmfloatwidth}
    \begin{center}
      \tmfloatcontents
      \captionof{#3}{#5}
    \end{center}
  \end{minipage}}
\begin{document}

\title{EPR-B correlations: \\
a physically tenable local-real model}\author{\and A. F. Kracklauer\\
Bauhaus University, Weimar, Germany}\maketitle

\begin{abstract}
  We propose a classical, i.e., local-real physical model of processes
  underlying EPR experiments. \ The model leads to the prediction, that the
  visibility of the output signal will exhibit increasing variation as the
  coincidence window is increased, thus providing a testable criteria for its
  validity. \ If it can be sustained, this model undermines the claim that
  Nature has a fundamentally nonlocal feature or that irreal entities are
  required by quantum theory.
\end{abstract}

\section{State of the argument.}

Historically, the struggle to to find an interpretation for the wave function
in quantum theory led, {\tmem{inter alia }}by way of arguments made by
Einstein, Podolski and Rosen (EPR), to examination of correlations of
measurements on systems comprising two entities. As is well known and commonly
accepted nowadays, analysis of such correlations seems to support the
conclusion that at a fundamental level, Nature admits either ``nonlocal,'' or
``irreal'' aspects \ The first of these alternatives constitutes a deep and
serious rift between the two main theories of Physics, namely Quantum
Mechanics and Relativity, because the latter demands ``locality,'' namely,
that interaction must transpire between all entities at or below the speed of
light. \ Accepting the alternative, irreality, is, if anything, an even deeper
break with occidental science, as it injects a role for human perception into
the evolution of the universe (via observer induced collapse of wave packets),
oblivious to the eons before humans appeared.

This situation has led some researches to critically reevaluate current
experiments deemed to justify the orthodox interpretations, in particular the
so-called EPR experiments and tests of `Bell-inequalities' as carried out by
Aspect, Weihs, etc. This writer, for example, has identified several generic
arguments supporting the conclusion, that the correlations seen in these
experiments do not result from structure unique to quantum mechanics, but
appear already in the analysis by Stokes of polarized light at least 50 years
before the need for quantum notions was recognized.{\cite{AK07}} \ Moreover,
this criticism of current thinking was under-girded with proposed simulations
of EPR experiments, intended to demonstrate in detail that non local
interaction is not needed to duplicate data commonly believed to require
quantum (i.e., nonlocal) interaction for explanation.{\cite{ak04}} \

Such simulations are incisive insofar as they constitute counterexamples to
the augment put forward by, for example, Bell, and subsequently dubbed a
``theorem,'' to the effect that EPR correlations absolutely cannot be taken
into account without the employ of nonlocality. \ Demonstrating a simulation
without nonlocality extinguishes that claim once and for all. \ Nevertheless,
these simulations can be, and have been, criticized for making use of features
that, although not quantum in their essential nature, are either impossible or
implausible as physics. \ This writer's simulation{\cite{ak04}} silently
assumes, for example, that photo detectors respond depending on the global
character of the input signal (specifically, which of the two component states
making up the `singlet state' actually enters the detector).{\cite{ga08}}

\ Less objectionably, the simulation proposed by de Raedt et al. proposes
that, upon deflection or transmission at a beam splitter, a ``photon'' either
suffers a random delay{\cite{sz08}}, or that beam splitters behave as
``deterministic learning machines,''{\cite{dR08}}. \ While such delays cannot
be rejected {\tmem{a priori}}, at least this writer knows of no physical cause
for such delays in beam splitters, in particular of the magnitude required by
the simulation to duplicate data taken in actual experiments. Alternatively,
``learning machine'' effects, as proposed, require, for example, persistent
polarization currents to be setup by one signal which persist to influence
subsequent signals passing through the beam splitter, an effect for which
explanation involves somewhat implausible, even extravagant hypothetical
input.

It is the purpose here to propose a physical model for these experiments that
accommodates the facts as observed in experiments. \ The model is simple,
fully classical, local and real; it does not presume any quantum structure,
not even the existence of ``photons.''

First, factual characteristics of the data from the experiments that are to
be modeled must be delineated.

\section{EPR-B data ``as it is.''}

\

The overwhelming impression made by EPR-B raw data, is that it appears to be
two (or four) streams of events occurring at essentially random
times{\footnote{See: {\cite{dR08}} for a comprehensive description of data
from a typical Bell-test experiment.}}. \ The two or four steams represent
what are called ``photon detections,'' at the detectors, either one (or two)
on both the right and left of an EPR setup. These data streams do not give the
impression to the eye of being temporally grouped or correlated. \ Thus, for
the purpose of analysis of the EPR experiments, these streams are filtered in
terms of a coincidence ``window,'' i.e., a time interval within which one
event on the left is paired with one event on the right. Naturally, as this
window is made more narrow, the number of resulting coincidences diminishes,
but at the same time the closer the statistics of the selected pairs come to
those predicted using quantum theory. At the other end, with a very wide
window, the coincidence statistics more closely approach those expected from
strictly classical analysis. \ The difference is, as is now well known, that
the quantum statistics involve more coincidences than are expected from non
quantum analysis. \ This is said to lead to ``quantum correlations stronger
than admitted by classical statistics.'' \

In addition, there is a second phenomenon revealed in the data. \ It is this:
the expected strict rotational invariance with respect to the input signals is
imperfect. \ This defect is visible as a variation in the visibility of the
output signal which oscillates so as to have minimums at the angles of $\pi /
4 + n \pi / 2$. \ To date, this anomaly remains under reported because it is
unexplained.{\cite{ak07s}}

Thus, what a simulation, or model, of these experiments must reflect and
explain goes beyond just the subset of the data which exhibit the peculiar
quantum aspects; it must also explain the large amount of data filtered out
which does not fully fit the quantum structure as revealed in calculations
with solutions to the Schr\"odinger equation{\tmname{{\tmsamp{}}}}. In this
regard, the two salient characteristics are:
\begin{enumeratenumeric}
  \item the relative paucity of qualified pairs, and
  
  \item the strict failure of rotational invariance as revealed by oscillating
  visibility.
\end{enumeratenumeric}
The model proposed below covers both of these features.

\section{The model: background motivation}

In a study of EPR and GHZ (after Greenburger, Horn and Zeilinger, who famously
considered experiments on higher order systems with three, four, etc. output
events) correlations, this writer has shown that the observed coincidences can
be accurately calculated using the {\tmem{classical {\tmem{variant of the
}}}}coherence function.{\cite{ak02}} For GHZ setups correlation calculations
become quite unwieldy, and, therefore, are best carried out with a computer
algebra program; here we use MuPAD. \ This tactic can be exploited also to
lead directly to a very convenient display of the essential difference between
the results of classical and quantum analysis of EPR (or GHZ if desired)
experiments (see below).

Consider a prototypical experiment to test Bell-inequalities. \ The
signal{\footnote{Herein the term ``signal'' refers to the pair of opposed
``pulses'' gnerated so as to have anti-correlated polarization. Each pulse is
taken to have just enough total energy to evoke one electron by the
photoelectric effect in a detector. \ As is customary, the ``signlet'' state
is considered to be a mysterious quantum entity comprised of the sum of two
such `signals' of opposite polarization orientation, which randomly
``collapses'' to one or the other component signal whenever one of the pulses
encounters a detector. }} as generated in the source crystal in the quantum
mechanical understanding of the setup constitutes a ``singlet state'' sent in
opposed directions, \ such that if a vertical signal is detected on the left,
a horizontal signal is detected deterministically on the right, or {\tmem{visa
verse}}. Such signlet states or signals emitted by the crystal are considered
to be essentially ambiguous in that they are the sum of both options in both
directions, but that this ambiguity is resolved by whichever detector
registers first, thereby causing a collapse of the quantum mechanical
$\psi$-function (wave) for the pulses sent in both directions. \ The
consequence is, that the mate to the first signal at the companion measuring
station is {\tmem{instantly}} granted (by some mysterious nonlocal process not
described by the dynamic equations of quantum mechanics) the opposite
polarization. \ In non quantum renditions or simulations of these experiments,
no such essential ambiguity is allowed; the paired signals are randomly
selected to be one or the other anti-correlated possibility and are never the
simultaneous sum in the form of the ``singlet state,'' \ a stipulation which
constitutes an encoding of ``reality'' in such simulations. A singlet state
must be considered ``irreal'' as it is supposedly the sum simultaneously of
mutually exclusive alternatives {\emdash} contrary to all logic.

For our model the calculation algorithm the source is encoded in terms of a
source signal sent left $\tmop{Sl} (n)$ and one sent right $\tmop{Sr} (n)$,
each a function of a random variable, $n$, taking on the values $0, 1$ with
equal probability. \ First, some initialization statements:

{\color{red}\ttfamily{{\color{red} $\bullet$ {\color{black} }}}}{\color{blue}
\verb|reset(): Matrix:=Dom::SquareMatrix(2): vector:=Dom::Matrix():|}

Now, the source signals are encoded as functions of $n$, such that for each of
its designated values, $0, 1$, it produces one or the other of the component
states constituting the singlet state :

{\color{red}\ttfamily{{\color{red} $\bullet$ {\color{black} }}}}{\color{blue}
\verb|Sl:=n->vector(2,1,[[n],[1-n]]); Sr:=n->vector(2,1,[[1-n],[n]]);|}

These signals are then each sent, on both sides, through polarizing beam
splitters (PBS). The axis of each beam splitter is separately variable, thus
the encoding of the effect of such a beam splitter on each beam must be a
function of its angular orientation. \ The operator corresponding to a PBS as
a function of its angular orientation is the 2-dimensional projection
operator:

{\color{red}\ttfamily{{\color{red} $\bullet$ {\color{black} }}}}{\color{blue}
\verb|proj:=z->Matrix([[cos(z),sin(z)],[-sin(z),cos(z)]]);|}

The output signal from a PBS is then obtained by multiplying the source signal
by the projection operator to obtain a two component vector, where each
component represents a signal from each output port of the PBS.

{\color{red}\ttfamily{{\color{red} $\bullet$ {\color{black} }}}}{\color{blue}
\verb|El:=(zl,n)->proj(zl)*Sl(n); Er:=(zr,n)->proj(zr)*Sr(n);|}

The output signals of the PBS, $(\tmop{El}, \tmop{Er}$), are electric fields
and are sent to photo-detectors, where they can be used in the standard law to
obtain the probability, $P$, of generating a photo electron, namely
\begin{equation}
  \text{$P \propto E^2,$}
\end{equation}
where $P$ can be interpreted as a photo current, for high intensity.

Now, whatever else is true or false, it is a fact, that the classical second
order correlation function corresponds to the same correlation as computed
with ostensible quantum algorithms. To execute this calculation we need to
form the ratio of difference over the sum of two other sums, of the form:
\begin{equation}
  \frac{\sum_{\tmop{ijkl}} \tmop{El}_i^{\ast} (\theta_l) \tmop{Er}_j^{\ast}
  (\theta_r) \tmop{Er}_k^{} (\theta_r) \tmop{El}_l (\theta_r) -
  \sum_{\tmop{ijkl}} \tmop{El}_i^{\ast} (\theta_l) \tmop{Er}_j^{\ast}
  (\theta_r + \pi / 2) \tmop{Er}_k^{} (\theta_r + \pi / 2) \tmop{El}_l
  (\theta_r)}{\sum_{\tmop{ijkl}} \tmop{El}_i^{\ast} (\theta_l)
  \tmop{Er}_j^{\ast} (\theta_r) \tmop{Er}_k^{} (\theta_r) \tmop{El}_l
  (\theta_r) + \sum_{\tmop{ijkl}} \tmop{El}_i^{\ast} (\theta_l)
  \tmop{Er}_j^{\ast} (\theta_r + \pi / 2) \tmop{Er}_k^{} (\theta_r + \pi / 2)
  \tmop{El}_l (\theta_r)},
\end{equation}
where all the indices take on the values $1$ \ and $2$ as the there are two
components representing the two output channels of a PBS.

In this expression one sees the definition of a forth order correlation of
electric field magnitudes, which for certain of these products equal second
order correlations of the field energy intensities, or, calling on the theory
of photo current generation, i.e., on the formula $I \propto E^2$, of the
correlation of the number of photo (electron) detections.

The encoded numerator of this expression is given by:

{\color{red}\ttfamily{{\color{red} $\bullet$ {\color{black} }}}}{\color{blue}
\verb|Num:=(n,c,i,j,k,l,zl,zr)->(-1)^c*El(zl,n)[i]*Er(zr+c*PI/2,n)[j]| \\
\phantom{xxxxx}\verb|*Er(zr+c*PI/2,n)[k]*El(zl,n)[l];|
}

{\ttfamily{$(n, c, i, j, k, l, \mathrm{\tmop{zl}}, \mathrm{\tmop{zr}}) \mapsto
\left( - 1 \right)^c  \mathrm{\tmop{El}} \left( \mathrm{\tmop{zl}}, n
\right)_i  \mathrm{\tmop{Er}} \left( \mathrm{\tmop{zr}} + \frac{\pi c}{2}, n
\right)_j  \mathrm{\tmop{Er}} \left( \mathrm{\tmop{zr}} + \frac{\pi c}{2}, n
\right)_k  \mathrm{\tmop{El}} \left( \mathrm{\tmop{zl}}, n \right)_l$}}

{\noindent}where the indicated sums are then executed by:

{\color{red}\ttfamily{{\color{red} $\bullet$ {\color{black} }}}}{\color{blue}
\verb|sum(sum(sum(sum(sum(sum(Num(n,c,i,j,i,j,zl,zr)| \\
\phantom{xxxxx}\verb|,i=1..2),j=1..2),k=1..2),l=1..2),n=0..1),c=0..1);|
}

{\ttfamily{$8 \left( \cos \left( \mathrm{\tmop{zl}} \right) \sin \left(
\mathrm{\tmop{zr}} \right) - \cos \left( \mathrm{\tmop{zr}} \right) \sin
\left( \mathrm{\tmop{zl}} \right) \right)^2 - 8 \left( \cos \left(
\frac{\pi}{2} + \mathrm{\tmop{zr}} \right) \sin \left( \mathrm{\tmop{zl}}
\right) - \sin \left( \frac{\pi}{2} + \mathrm{\tmop{zr}} \right) \cos \left(
\mathrm{\tmop{zl}} \right) \right)^2$}}

{\color{red}\ttfamily{{\color{red} $\bullet$ {\color{black} }}}}{\color{blue}
\verb|simplify(%);|}

{\ttfamily{$- 8 \cos \left( 2 \mathrm{\tmop{zl}} - 2 \mathrm{\tmop{zr}}
\right)$.}}

The denominator, $\sum_{\tmop{ijkl}} \tmop{El}_i^{\ast} (\theta_l)
\tmop{Er}_j^{\ast} (\theta_r) \tmop{Er}_k^{} (\theta_r) \tmop{El}_l (\theta_r)
+ \sum_{\tmop{ijkl}} \tmop{El}_i^{\ast} (\theta_l) \tmop{Er}_j^{\ast}
(\theta_r + \pi / 2) \tmop{Er}_k^{} (\theta_r + \pi / 2) \tmop{El}_l
(\theta_r)$, is computed with the same statements with exception of the factor
of $(- 1)^c$, which is replaced with $(+ 1)^c$:

{\color{red}\ttfamily{{\color{red} $\bullet$ {\color{black} }}}}{\color{blue}
\verb|Den:=(n,c,i,j,k,l,zl,zr)->(+1)^c*El(zl,n)[i]*Er(zr-c*PI/2,n)[j]| \\
\phantom{xxxxx}\verb|* Er(zr-c*PI/2,n)[k]*El(zl,n)[l]:|
}

{\color{red}\ttfamily{{\color{red} $\bullet$ {\color{black} }}}}{\color{blue}
\verb|sum(sum(sum(sum(sum(sum(Den(n,c,i,j,i,j,zl,zr)| \\
\phantom{xxxxx}\verb|,i=1..2),j=1..2),k=1..2),l=1..2),n=0..1),c=0..1): simplify(%);|
}

{\ttfamily{$8$.}}

Thus, this calculation delivers the same result as does the quantum algorithm
for the correlation. Like many such quantum algorithms, the physical
interpretation is less than obvious. \ But, for present purposes, an
interesting point of entry is the numerator of the correlation. \ The result
obtained above, namely
\begin{equation}
  \left( \cos \left( \mathrm{\tmop{zl}} \right) \sin \left( \mathrm{\tmop{zr}}
  \right) - \cos \left( \mathrm{\tmop{zr}} \right) \sin \left(
  \mathrm{\tmop{zl}} \right) \right)^2 - 8 \left( \cos \left( \frac{\pi}{2} +
  \mathrm{\tmop{zr}} \right) \sin \left( \mathrm{\tmop{zl}} \right) - \sin
  \left( \frac{\pi}{2} + \mathrm{\tmop{zr}} \right) \cos \left(
  \mathrm{\tmop{zl}} \right) \right)^2,
\end{equation}
offers a propitious venue for interpretation of the physics involved by
expanding each term, i.e.,
\begin{equation}
  \cos^2 \left( \mathrm{\tmop{zl}} \right) \sin^2 \left( \mathrm{\tmop{zr}}
  \right) - 2 \cos^{} \left( \mathrm{\tmop{zl}} \right) \sin^{} \left(
  \mathrm{\tmop{zr}} \right) \cos \left( \mathrm{\tmop{zr}} \right) \sin
  \left( \mathrm{\tmop{zl}} \right) + \cos^2 \left( \mathrm{\tmop{zr}} \right)
  \sin^2 \left( \mathrm{\tmop{zl}} \right) + \Upsilon,
\end{equation}
where $\Upsilon$ is the expansion of the second term in Eq. (3). The first and
third terms of Eq. (4) are the standard expressions for correlations of events
on each side from each of the variant signals put out by the source; but, the
middle term is not standard. \ It does not fit the law of photo current
generation, i.e., $P \propto E^2$; because it is the product from four
different fields, giving the amplitudes of electric fields for four different
photo electron source events {\emdash} all of which appears to mean that it
cannot be associated with the generation of a photo electron current. \

In short, it is this middle term that requires a model or interpretation if
the mystical aspects of EPR correlations are to be explained without recourse
to preternatural phenomena.

\section{A new model for EPR correlations}

The motivation for the model proposed here is provided by the most conspicuous
feature of that data from EPR experiments data, namely, that data exhibiting
the so-called quantum correlations is just a subset of the total data steam
which has been filtered out in terms of a ``coincidence window.'' The
narrower, the window, the closer the coincidence pattern approaches the ideal
as calculated using the quantum algorithm.

This fact suggests the following structure, which shall be taken as the
hypothetical input for our model. The source crystal under stimulation of the
driving input beam at various emission centers in the crystal independently
produces two types of anti corrected output signals, one with a vertically
polarized pulse to the right and a horizontally polarized pulse to the left,
and one in which the signal has switched polarization orientations. \ It is
taken that the separate pulses in each pair have an intensity just sufficient
(statistically) to elicit one photo electron.{\footnote{The absolute
anti-correlation of detections in the output channels of a a beam splitter at
the one photo electron level is considered nowadays to reflect quantum
structure. However, behaviour at a PBS is irrelevant to the quantum structure
of coincidences. \ Moreover, there are also non quantum models for beam
splitters, e.g., {\cite{es05}}, that can purge this last vestiges of quantum
structure from the analysis of EPR correlations.}} \ \ Nevertheless, it is not
taken that the matched pulses, one in each arm, necessarily elicit photo
electrons exactly simultaneously; as is well known they can be separated by
some random time interval, compatible with the coherence length of the pulses.
Now, filtering the data to fit within a ``window,'' selects those coincidences
of one detection on each side engendered by these two types of signal pairs
that happen to have evoked photo electrons coincidentally within the window.
Such coincidences arise from two distinct situations: one, when each of the \
photo electron detections is \ elicited by the same signal, and, two, when an
event on each side is elicited by oppositely polarized signals. \ The first
sort correspond to the first and third terms in Eq. (4).

But, there are still additional coincidences possible if two separate signal
types from the crystal are timed such that a coincidence can arise between
single detections evoked one by each signal. \ In terms of classical physics,
this assumes that two such signals, one of each variant, arise at separate
locations within the crystal. Coincidences of this sort, which can be
considered ``illegitimate,'' obviously, can be involved only if these two
coincidences fall within a very short interval, shorter than the expected
interval between coincidences generated by ``legitimate'' pairs, thereby
forming a closer time association than the underlying legitimate events. When
the window is narrow these illegitimate coincidences are counted along with a
certain number of legitimate ones, whereas, with a broad window, additional
coincidences are counted which arise only from legitimate pairs, i.e., from
signals that are not coincident with the opposite variant. These extra,
legitimate counts dilute the selected data sample by adding relatively more
coincidences not corresponding to the cross terms in Eq. (4).

This can be understood in terms of the probabilities of detection. \ For each
detection of a photo electron resulting from a pulse (assumed to have energy
sufficient for only one photo electron) the probability that this electron is
lifted into the conduction band after a time $\tmop{dt}$ equals $\tmop{dt} /
l$ where $l$is the pulse length. The probability of a coincidence with the
lifting of a photo electron in the companion measuring station is then equal
to $(\tmop{dt} / l)^2$, i.e., the probability of the coincidence of the two
events is the product of their probabilities. \ Now, for the coincidence of
photo electrons engendered by ``illegitimate'' combinations, the probability
equals $(\tmop{dt}_1 / l) (\tmop{dt}_2 / l)$ where the subscripts indicate
that two distinct signals are under consideration, i.e., an ``illegitimate''
coincidence. Clearly, to conform to the calculation described above, which is
fully symmetric in the admitted combinations (as encoded by selection of
indices) over which the sum is carried out, these two probabilities must be
equal. \ This can arise only when $\tmop{dt}_1 \equiv \tmop{dt}_2$, that is,
when the two signals from the crystal are by chance simultaneous because then
the two intervals will be equal, or nearly so, only when they have essentially
identical initial instants, which applies tautologically for coincidences that
are virtually simultaneous i.e., when selected under a very narrow coincidence
window. \ However, for illegitimate coincidences, there is likely an offset in
the starting instants for the $\tmop{dt}_i$ \ as they pertain to separate,
uncoordinated, independent signal pairs. \ The extent to which the data
includes coincidences admitted by a wide filter, it the extent to which the
quantum pattern is corrupted. \ \ \

A central issue is: can these `illegitimate' coincidences be related to the
cross terms in Eq. (4)?

At first glance, this would seem to be impossible because the form of this
term does not admit a straightforward interpretation in terms of the law of
photo electron generation, which requires electric field amplitudes squared. \
Just here, however, another possibility enters. \ It is this. \ If it is
allowed, that coincidences can arise from the occasions on which there is a
temporal overlap of elementary signals, one of each variant, then there is a
formal equivalence between the number of additional coincidences and the cross
terms in Eq. (4). \ The essence of this equivalence consists in the fact, that
formally seen, the cross terms in Eq. (4) correspond to physically sensible
terms of the form:
\begin{equation}
  E_i (\tmop{zl})^2 E_j (\tmop{zr})^2,
\end{equation}
obtained with the substitutions:
\begin{equation}
  \cos (x + \pi / 2) = - \sin (x) ; \sin (x + \pi / 2) = \cos (x),
\end{equation}
that convert the four-factor, cross terms in Eq. (3) to the form of Eq. (5),
which does admit physical interpretation in terms of photo electron
generation. The phase shifts of $\pi / 2$ in the arguments here are related to
the fact that the two factors on each side, pertain to opposite variants of
the source signals, which must take into account the difference between the
polarization orientation of the input pulses entering the beam splitters. \

Physically, this correspondence take coincidences into account which arise
from a photo electron generated on the left from one of the variants with a
detection on the right from the opposite variant which happens to overlap
temporally. \ The number of such coincidences corresponds to the number
arising from source signals comprised of correlated (in stead of
anti-correlated) pairs. \ This poses a challenge for a physical interpretation
of these experiments because the phase matching conditions imposed on the
nonlinear generation processes in the crystal produces only anti-correlated
output pairs. \ This challenge (for formulating a model) \ is overcome,
however, by positing that such correlated pairs are comprised of detections on
each side matched with detections on the other side from a distinct signal of
the opposite variant.

\ It is to be stressed, that the significance of Eqs. (6) is {\tmem{not
{\tmem{}}}}strictly physical. \ It represents mostly a formal correspondence
rendering the quantum algorithm in accord with a feasible physical process.

In other words, detections are considerably different if two different but
overlapping signals are present. \ Then, it becomes possible, that the two
photo electrons closest in time have arisen from different pairs. \ That
means, that in the total population of detected pairs, as identified by
proximity of detection times, among pairs selected by narrow window criteria,
there will be some pairs formed corresponding to the cross terms in Eq. (4). \
But as the window width is increased, so as to capture additional pairs, the
number of such illegitimate pairs is not greater than the number found using a
narrow window. \ In other words, the total data set grows in size with
increasing window width but the number of illegitimate pairs or pairs
corresponding to the cross terms remains the same. \ This explains just why
experimenters must strive to reduce the window width as much as feasible; and,
it explains why only the statistics of a subset of data taken with a narrow
coincidence window width conform with those computed using so-called quantum
algorithms.

\section{The rotational invariance anomaly}

A crucial feature of EPR data to be explained by any model is the evidence of
a breakdown of rotational invariance. \ Previous analysis of this matter by
this writer led to the conclusion that this break down should be complete,
i.e., that the visibility of the signal should actually approach null for
certain angular settings of the polarizers in the detection
stations.{\cite{ak07s}} This conclusion, however, resulted from the assumption
that individual signals are exclusively of one variant or the other. \ As
described above, in the model proposed herein, such pure signals are indeed
the majority, except for that small subset of data filtered out when the
detection window width approaches null. \ These selected events in this
special case comprise events corresponding to the cross terms from the overlay
of both variants of source signals, and, as such, under optimum conditions,
constitute a rotationally independent sum. \ Each signal variant under
rotation compensates the other as they are exactly out of phase. \ This means
that the small variations in visibility still seen in the data result from the
fact that the window width is still sufficiently wide to admit a less than an
ideal set of coincidences, that is, it contains a relative deficit of cross
terms so that the statistics of this data set exhibit some portion of
rotational variance.

This effect can be quantitatively depicted by redefining Eq. (4) so that the
legitimate terms are multiplied by a factor, which when greater than $1$,
represents the excess in their number whenever the coincidence window is
widened:

{\color{red}\ttfamily{{\color{red} $\bullet$ {\color{black} }}}}{\color{blue}
\verb|cor:=(zl,zr,y)->-cos(zl)^2*cos(zr)^2+cos(zl)^2*sin(zr)^2 | \\ 
\phantom{xxxxxxxxxx}\verb|- y^-1* 4*cos(zl)*cos(zr)*sin(zl)*sin(zr)|  \\
\phantom{xxxxxxxxxxxxxxx}\verb|+cos(zr)^2*sin(zl)^2-sin(zl)^2*sin(zr)^2;|
}

{\ttfamily{$( \mathrm{\tmop{zl}}, \mathrm{\tmop{zr}}, y) \mapsto \left( \left(
\cos \left( \mathrm{\tmop{zl}} \right)^2 \sin \left( \mathrm{\tmop{zr}}
\right)^2 - \cos \left( \mathrm{\tmop{zl}} \right)^2 \cos \left(
\mathrm{\tmop{zr}} \right)^2 \right) - \frac{4 \cos \left( \mathrm{\tmop{zl}}
\right) \cos \left( \mathrm{\tmop{zr}} \right) \sin \left( \mathrm{\tmop{zl}}
\right) \sin \left( \mathrm{\tmop{zr}} \right)}{y} \right) + \cos \left(
\mathrm{\tmop{zr}} \right)^2 \sin \left( \mathrm{\tmop{zl}} \right)^2 - \sin
\left( \mathrm{\tmop{zl}} \right)^2 \sin \left( \mathrm{\tmop{zr}}
\right)^2$}}

{\noindent}This expression is the result of multiplying all terms for
legitimate coincidences in both the numerator and denominator and is therefore
the absolute correlation function as a function of $y$, the excess factor of
legitimate coincidences resulting from non ideal filtering. \ By now
substituting this into the expression for CHSH discriminator:

{\color{red}\ttfamily{{\color{red} $\bullet$ {\color{black} }}}}{\color{blue}
\verb|S:=(x,xx,z,zz,y)->cor(x,z,y)-cor(x,zz,y)+cor(xx,z,y)+cor(xx,zz,y):|}

{\noindent}and then defining, for convenience, the function:

{\color{red}\ttfamily{{\color{red} $\bullet$ {\color{black} }}}}{\color{blue}
\verb|SS:=(w,v,y)->S(w,w+2*v,w+v,w+3*v,y);|}

{\noindent}which is defined in terms of the values of the orientations of the
polarizers which have been determined to maximise the violation of a
Bell-inequality. \ $\tmop{SS} (w, v, y)$ can be plotted directly; see figures.

%\tmfloat{h}{small}{figure}{\resizebox{2in}{!}{\includegraphics[scale=3]{fig_1.eps}}}{}\tmfloat{h}{small}{figure}{\resizebox{2in}{!}{\includegraphics[scale=3]{fig_1.1.eps}}}{}
%\tmfloat{h}{small}{figure}{{\includegraphics[scale=0.4]{fig_1.eps}}}{}\tmfloat{h}{small}{figure}{{\includegraphics[scale=0.4]{fig_1.1.eps}}}{}

\begin{figure}\noindent
\begin{minipage}[b]{.5\linewidth}\centering\epsfig{figure=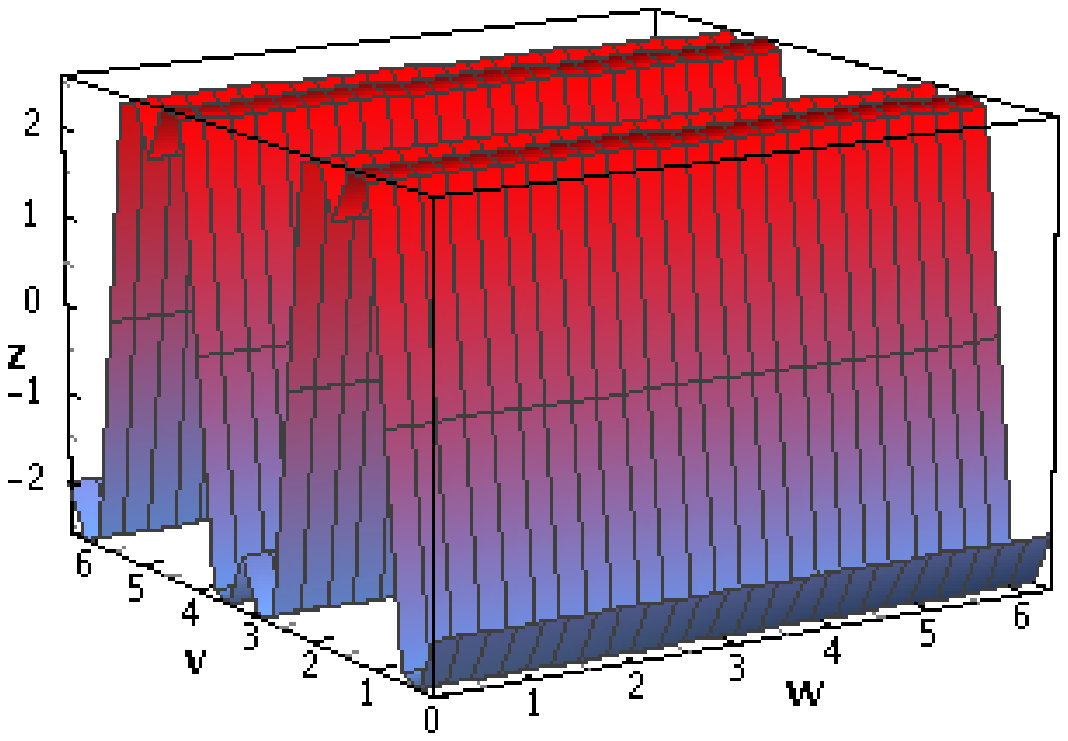,width=\linewidth}\caption{Ideal data set}\end{minipage}
\begin{minipage}[b]{.5\linewidth}\centering\epsfig{figure=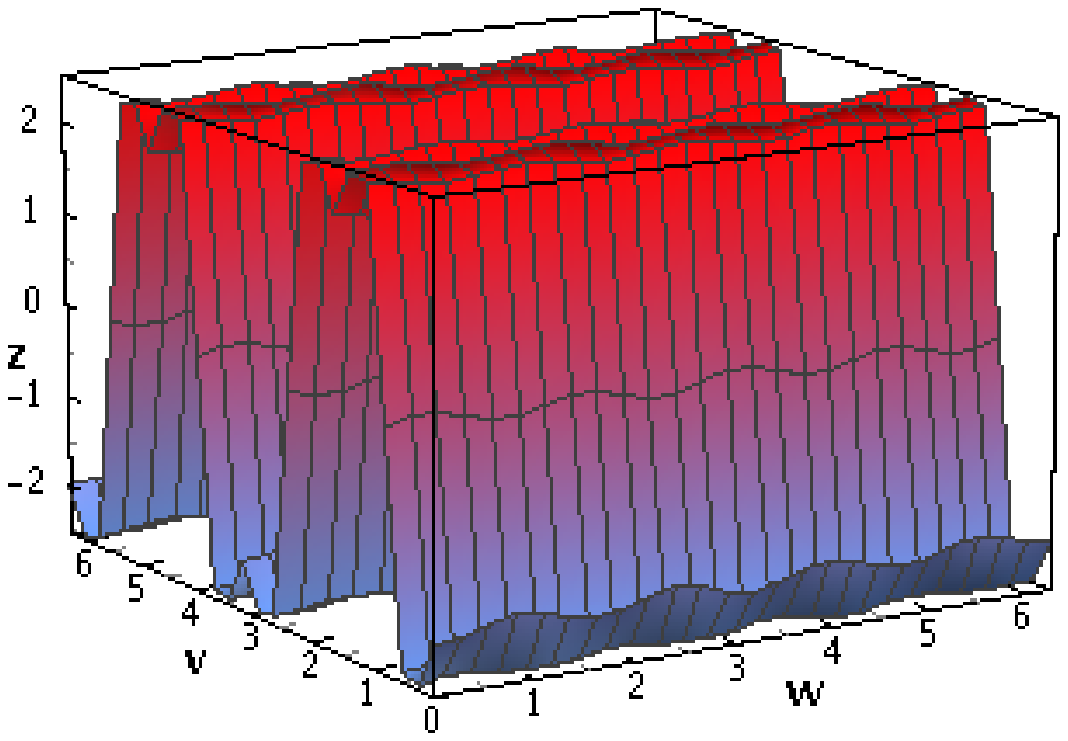,width=\linewidth}\caption{With excees legitemate coincidences}\end{minipage}
\end{figure}

Fig. 1 shows the variation of the function $S$ in terms the variables $v$, the
displacement between the angular settings on both sides in the various
experiments contributing terms to the CHSH discriminator when $y = 1$, i.e.,
for an ideal subset of data. If the source signals are rotationally invariant,
then the values for $v$ have no absolute meaning, which implies that variation
of $w$, the starting point, has no effect on $S$. \ Fig. 2 illustrates just how much rotational
invariance is destroyed by a 10\% increase in the number of legitimate
coincidences in the total data stream.

The existence of this effect is strong evidence that the source signals are
not, as envisioned in orthodox quantum theory, comprised of singlet states;
rather, the singlet state structure is mimicked only in the (ideal) subset of
data filtered out with a small coincidence window.

\section{Conclusions} \ \

In the introduction characteristics of the data taken in EPR experiments that
a physical model of the underlying physical phenomena must explain were
identified. \ It was noted that the data taken in these experiments differs
considerable from the ideal data predicted by conventional analysis based on
current quantum orthodoxy. \ A viable physical model should explain all the
data taken, that is, the empirical facts, rather than just a sub portion,
however significant it may be considered.

The three salient features of the data are explained by the model as follows:

\begin{enumeratenumeric}
  \item {\tmem{The most obvious feature of EPR data streams is that there are
  very few obvious, ideal coincidences}}.
  
  The proposed model incorporates or explains this feature as a consequence of
  the distribution of emission time of photo electrons. In this case it is
  taken that the emission time of the photo electron in a detector is
  displaced naturally and randomly within a pulse length or coherence length
  of the signal impinging on the detector. \ It is a well know fact that
  electromagnetic pulses, while they can stimulate virtually instantaneous
  emission of photo electrons, in fact for an ensemble of such pulses, the
  actual exact emission times are distributed over the pulse length. Thus, the
  emission times on opposite sides of an EPR setup usually do not exactly
  coincide, even when generated by the same signal or pulse pair.
  
  \item {\tmem{The population of coincidences that exhibit the pattern as
  calculated using algorithms from quantum mechanics must be filtered from the
  total data set by selection using ``coincidence circuitry'' with as narrow a
  time interval window as possible}}.
  
  Data subsets selected with a wider than optimum window, while they may
  exceed limits considered to obtain from non quantum regimes, nevertheless
  deviate from the ideal quantum pattern and approach that for non quantum
  systems. \ Specifically, the curve of the CHSH discriminator function
  exceeds $2$ but is significantly less that $2 \sqrt{2}$.
  
  In terms of the proposed model, these features are explained as follows: \
  It is taken that occasionally two pairs of signals of opposite polarization
  character overlap temporally. \ For theses cases, it might happen that the
  closest coincidences involve detections from each signal because the actual
  displacement of the truly paired events (i.e., from the same signal) are
  displaced at larger intervals by cause of the effect considered in point 1.
  above than these ``illegitimate'' coincidences. These extra, illegitimate,
  coincidences are the excess that correspond to the cross terms in Eq. (3),
  so that the manipulations in the quantum algorithm yield the observed
  patterns coincidentally, even while these algebraic expressions as such do
  not correspond to direct application of known physics principles, here the
  photoelectric effect.
  
  The corruption of the data set as the coincidence window is broadened is
  then a consequence of the fact, that the ``illegitimate'' coincidences
  represent relatively rare events that cannot occur for well separated signal
  variants. They occur only for nearly simultaneous sums of both variants.
  Therefore as the window is increased, the set of events considered for
  analysis includes ever increasing numbers of coincidences formed from
  legitimate pairs, and the statistics approach those for classical
  particulate systems.
  
  \item {\tmem{The filtered data sets for even narrow windows exhibits a
  variation in visibility that should not occur for the data set envisioned in
  terms of quantum analysis (i.e., for the singlet state, which is perfectly
  rotationally invariant).}}
  
  This phenomenon in terms of the model proposed herein is a result of the
  fact, that no matter how narrow the coincidence window is taken, the
  statistics can only approach the ideal which is mathematically codified with
  the singlet state. \ If this model is faithful to the actual physical facts
  of the EPR experiments, then it is to be expected that the degree of
  violation of rotational invariance, and therefore the visibility, is a
  function of the window width {\emdash} a testable proposition.
\end{enumeratenumeric}
If this model is empirically sustainable, then it has been shown, that claims
that EPR experiments and experimental tests of ``Bell's theorem'' prove that
Nature at a fundamental level involves nonlocal interaction, cannot be
maintained. \ Likewise, it would show, that ``irreal'' states (i.e., states
composed of the sum of mutually exclusive options, for example the singlet
state) do not have ontological status. \ They are artifacts of the formalism
for which statistical parameters are valid for ensembles, but that cannot be
applied to the individual entities constituting the ensemble. This conclusion
follows directly from the fact that, were the input variant signals actually
singlet states in each individual case, the statistics would be essentially
invariant with respect to the coincidence window width because all signals
would have equal probability of producing ``illegitimate'' coincidences.

Of course, it is in principle possible that some other true ``quantum'' system
in which singlet states arise in fact exists. \ Experiments on such a system
might then verify the current understanding of the issues around nonlocality;
but, until such experiments on such systems are carried out, these issues
remain open.

\end{document}